\documentclass[12pt]{article}
\usepackage{graphicx} 
\usepackage{bbm}
\usepackage{epsfig,amssymb,latexsym,color,amsmath}       
\DeclareGraphicsExtensions{.pdf}

\newcommand{\n}{\noindent} 
\newcommand{\C}{\Sigma}
\newcommand{\be}{\begin{equation}}
\newcommand{\ee}{\end{equation}}

\begin{document}

\title{\textbf{Modeling crowdsourcing as collective problem solving}}
\author{A. Guazzini${}^{1,2}$, D. Vilone${}^{3}$, C. Donati${}^{1}$, A. Nardi${}^{1}$, Z. Levnaji\'c${}^{4,5}$}
\date{}
\maketitle

\n ${}^{1}$\textit{Department of Science of Education and Psychology, University of Florence, Florence, Italy} \\ 
${}^{2}$\textit{Center for the Study of Complex Dynamics, University of Florence, Florence, Italy} \\
${}^{3}$\textit{Laboratory of Agent Based Social Simulation (LABSS), Institute of Cognitive Science and Technology, National Research Council, Rome, Italy}  \\
${}^{4}$\textit{Faculty of Information Studies in Novo mesto, Novo mesto, Slovenia} \\
${}^{5}$\textit{Faculty of Computer and Information Science, University of Ljubljana, Ljubljana, Slovenia}

\section*{Abstract}  \label{Abstract}
\n Crowdsourcing is a process of accumulating the ideas, thoughts or information from many independent participants, with aim to find the best solution for a given challenge. Modern information technologies allow for massive number of subjects to be involved in a more or less spontaneous way. Still, the full potentials of crowdsourcing are yet to be reached. We introduce a modeling framework through which we study the effectiveness of crowdsourcing in relation to the level of collectivism in facing the problem. Our findings reveal an intricate relationship between the number of participants and the difficulty of the problem, indicating the optimal size of the crowdsourced group. We discuss our results in the context of modern utilization of crowdsourcing.

\

\

\

\noindent {\scshape Reference:} {\it Scientific Reports}, {\bf 5}, 16557 (2015).


\thispagestyle{empty}

\section*{Introduction} \label{Introduction}

We learn by solving problems. During evolution humans have developed the ability not only to learn individually, but also to share the acquired knowledge. Indeed, there is little doubt about the role played by the natural selection in the emerging social nature of humans~\cite{darwin,dunbar}. Human beings probably have an innate tendency to operate in groups: \textit{Pan troglodytes} (common chimpanzee), the closest living relative of modern humans, is known to have a noticeable social structure~\cite{janegoodall}. The emerging skill of social problem solving has allowed our ancestors to face the challenges of gradually increasing complexity. This factor not only determined our evolutionary success, but it still shapes our cultural and social behavior~\cite{moleon}. In addition, everybody has at least minimal capacity to coordinate the activity of others and direct the process of problem solving~\cite{dumas,jiang}. However, the existence of this ability is by no means obvious, as it requires an intricate synchronization between perception, communication and cognitive representation of the situation-task-others triad. Social Darwinism provides a framework to study the evolution this capacity, not just in terms of neurophysiology, but also in relation to the cognitive activities behind social problem solving.

On the other hand, any activity of intrinsically social nature requires a robust level of collaboration between the interacting subjects. The existence of a large sub-population of cooperators guarantees a stable support for the community, but at the same time creates a socioecological \textit{niche} for the ``free-riders'' -- the subjects benefiting from the available social support while providing little to its development. The free-riders are in fact known to be a well identifiable sub-population, competing to survive within various societies~\cite{baumol,cornes}. Still, their individualism in facing new tasks can eventually create new knowledge that could not have been created by working only as a collective. An individual facing a new challenge in isolation has considerably smaller chances of solving it, despite relying on the available knowledge. Yet, if a solution under these circumstances is ultimately found, this subject learns much more than when the solution is found collectively. Modeling this phenomenon is at the hearth of the paper that follows.

Fast forward to the present day, engaging citizens and stakeholders in gathering constructive ideas and practices is among the most interesting challenges of our time. Referred to as \textit{crowdsourcing}~\cite{howe,surowiecki}, this process is often defined as a ``collective intelligence system''~\cite{arolas,zhaozhu}. Crowdsourcing for ideas or solutions means that many subjects freely contribute independent suggestions. If a functional system of managing these suggestions is established, crowdsourcing can lead to surprising results, in the best sense of collective problem solving. In his 2009 Polymath project, Cambridge mathematician Tim Gowers proposed a very hard problem in his blog, that he estimated would take him months to solve. In only a few weeks, the community of his colleagues solved not only the original problem, but also a far much harder generalized version of it~\cite{polymath}. Needless to say, Wikipedia is probably the most successful example of the applied crowdsourcing~\cite{taha1,gandica}. Indeed, the extraordinary development of information and communication technologies calls for new practices of ``virtual participation'', that would enable the simultaneous engagement of a huge number of individuals~\cite{behrend}. Diverse studies and applications of crowdsourcing include crowd science~\cite{lee,sauermanna}, public participation in planning projects~\cite{brabham1}, public health~\cite{brabham2} and crowd funding~\cite{prpic}. Still, the full potentials of crowdsourcing in the digital era remains underused. Social media is probably the most powerful platform for this, which however requires a mechanism for systematic management of users' engagement. In this context, the precise understanding of the origins and roles of free-riders in crowdsourcing has still to be strengthened, which is the core focus of this paper.

Psychologists have repeatedly shown that certain general intelligence

\noindent emerges from the correlations among peoples' performance~\cite{kearns,judd}. But the true nature of ``collective intelligence'', which is expected to determine the capacity of collective problem solving, remains elusive~\cite{hackman,sornette}. Woolley et al. found a converging evidence of collective intelligence by explaining a group performance on a wide variety of tasks~\cite{woolley}. Upon  establishing the name ``$c$-factor'' for the obtained value of collective intelligence, they fund this value not to be strongly correlated with the average or maximum individual intelligence of the group members, but instead with other parameters such as social sensitivity, conversational turn-taking and the proportion of females. On a different front, there are massive efforts to develop computational~\cite{matjaz1,matjaz2,matjaz3} and experimental models~\cite{jela1,bond,jela2,phana} of collective social behavior, primarily within the framework of game theory and computational social science~\cite{easley,conte,torok}. These studies systematically show that cooperation is in general always beneficial for the community, which calls for mechanisms to discourage defection and oppose its strong appeal as a temporary benefit for an individual~\cite{mason,daniele,misvi,rand2,andrea1,matjaz4}.

All these findings reveal an extremely complex scenario, in which too many dimensions must be taken into account to obtain even a nearly realistic model of crowdsourcing. Seeking to implement at least the core assumptions, we propose a stylized model linking the process of crowdsourcing to better studied process of collective problem solving. Our model is grounded in the fact that a (social) player can choose to either share the acquire knowledge with others or not, which in turn determines the amout of help obtained from the community, dictating the improvement of this player’s individual skills. Our model is implemented both computationally as a stochastic evolutionary system, and analytically via methods of statistical physics. The model is inspired by the sociopsychological bases of knowledge dynamics that follow from the discussion above. As we show, similarly to defection emergence in game-theoretic models, free-riding here is a direct consequence of the evolutionary nature of our model and the interaction between the involved degrees of freedom. However, despite evolving towards maximizing the knowledge of the individual subjects, our model reveals a critical group size where an interesting balance between collective and individual approaches is maintained. This is, to our best knowledge, an entirely new model of crowdsourcing.

\section*{Model} \label{Model}

Below we present the computational model, which for better clarity, we do in several steps. First we explain the process that we wish to model. Our interest is to see how dividing a given population with fixed number of subjects (below called players) into several smaller groups, will influence the ability of these groups to cope with the problems (below called tasks) of different difficulty. We look on one end at the situation where everybody is in the same group, and on the other at the scenario of each player working alone. Between these two extremes, we look for the optimal group size that would allow for its players to learn the most. We are also interested in how the average intensity of collaboration relates to the group size. \\

\n \textbf{General setting.} We consider a population of $N$ players divided into $n$ groups. Each group has the same size (number of players) $S$, so that $S=\frac{N}{n}$. We take $N=100$, and consider seven specific choices of $S$ to keep $n$ integer. These sizes are $S=1,2,5,10,25,50,100$, respectively corresponding to dividing the population into $n=100,50,20,10,4,2,1$ groups. The total number of players $N=100$ always remains fixed. \\

\n \textbf{Collaboration and task assignment.} Once $n$ and $S$ are selected, we proceed by assigning a value $p_i$ to each player $i$ in each group. Each $p_i$ is chosen uniformly at random between 0 and 1. $p_i$ identifies the player's tendency to collaborate with other group members when solving a certain task (the tendency to work collectively as opposed to individually). Small values of $p_i$ (close to 0) indicate a preference for individualism, while large values (close to 1) indicate a tendency to collaborate. The values $p_i$ do not change and are hence characteristic for each player. Next, a group is assigned a task to be solved, quantified by the value $R$, which is randomly uniformly chosen between 0 and 1. $R$ measures the simplicity of the task, in the sense that $R$ close to 1 indicates an easy task, whereas $R$ close to 0 implies a hard task (in other words, the difficulty of the task is $1-R$). Each of $n$ groups that work in parallel is assigned the tasks of the same simplicity $R$. For $R$ we consider the following thirteen choices $R=0.0001, 0.001, 0.01, 0.1, 0.2, 0.3, 0.4, 0.5, 0.6, 0.7, 0.8, 0.9, 1$. Note that our scale for smaller $R$ is exponential, while it is linear for larger $R$, in order to better examine very hard tasks where collective working is expected to be beneficial. \\

\n \textbf{The definition of a simulation.} For each choice of group size $S$, and for each choice of tasks simplicity $R$ (both among the pre-specified $7 \times 13$ values mentioned above), we run a sequence of crowdsourcing games detailed below. Thus, each simulation is carried out for a fixed choice of $S$ and $R$, which are the two main parameters in our analysis that follows. Our interest is to examine the task solving behavior of a group of $S$ players when confronted with a task of simplicity $R$. The games in each simulation are done separately within each of the $n$ groups. \\

\n \textbf{One iteration.} One iteration (or a time-step) for a given group consists of the following three steps:
\begin{itemize}
 \item We establish a subgroup of players which play this iteration together and call them 'collectivists'. Each player $i$ joins the collectivists with the probability $p_i$. Each player that did not join the collectivists, plays this iteration alone. We call each such player an 'individualist'. Thus, a group is in each iteration divided into a subgroup of collectivists, in addition to the remaining number of individualists. Thus, larger $p_i$ means that for most iterations this player will end up as a collectivist.
 \item For each individualist separately we choose a number for this iteration $G_i$. This $G_i$ is the desired gain for this player, whose meaning is clarified below. $G_i$ is an integer chosen uniformly at random between 1 and 10. Larger $G_i$ means smaller chance to win (solve the task), but with potentially bigger gain if the task is actually solved. 
 \item Now, each collectivist solves the task with the probability $R$, while each individualist solves the task with the probability $R^{G_i}$ (since $R \le 1$, larger $G_i$ means smaller chance of solving). These two different probabilities model the fact that by approaching the task in isolation, an individualist has in fact to solve a $G_i$ times harder task. In contrast, collectivists face the task together which improves the chance for each of them to solve it. This difference is reflected in their respective scores, as explained below.
\end{itemize}
Each iteration finishes with some players winning (solving the task) and others losing (not solving the task). Both a collectivist and an individualist can win or lose. Next iteration begins with a new draw of collectivists using the same values $p_i$. These rules apply equally for all $n$ groups that play simultaneously and in all iterations. \\

\n \textbf{Group's capacity and player's fitness.} Once the winners and losers for a given iteration are known, we proceed with assigning the scores. For a given group we introduce an integer parameter $\C_j$ called capacity (we use $j$ to index groups, in order to differentiate from $i$ which indexes players within a group). Capacity is simply the number of iterations in which one collectivist has solved the task, regardless of $R$. It models the collective knowledge of a group, i.e., the group's capacity to solve increasingly challenging tasks. It is defined for each group, regardless of its iteration-dependent divisions into collectivists and individualists. Initially we set for all groups $\C_j = 0$. We update $\C_j \rightarrow \C_j + 1$ each time one player playing as a collectivist solves the task. Next we introduce for each player a parameter called fitness or pay-off $\pi_i$. It models player's own benefit in terms of new knowledge acquired. A collectivist who solves a task contributes to each player in the group (collectivists and individualists) the additional fitness $\frac{\C_i}{S}$, where $\C_i$ is the updated capacity. A collectivist who fails to solve the task does not contribute anything. An individualist who solves the task gains additional fitness $G_i$ without contributing anything to the other players, while an individualist who fails makes no additional gain. Still, each individualist always gains $\frac{\C_i}{S}$ per each collectivist that had solved the task. This organization of pay-offs models the fact that collectivists distribute new knowledge both to themselves and to all others, while individualists keep it for themselves. Collectivists however solve easier tasks since they work together, but with potentially lesser new knowledge (fitness) gained for each of them separately. In contrast, by working alone, individualists solve harder tasks but learn much more when they actually solve them, while avoiding to share this new knowledge with anyone else. \\

\n \textbf{Motivation behind two scoring scales.} Our model attempts to portray the incremental nature of human advances, for which there is evidence of superlinear behavior. To illustrate with a toy example, the invention of the microscope (first advancement), linked to the discovery of a new molecule (second advancement), can ultimately lead to the creation of the cure for a dangerous desease (third advancement). This represents a chain of fitness gains, with the total gain much larger than the simple addition of the payoffs due to single advancements in optics and chemistry. Thus, two distinct gaining schemes can be established: the individual scheme related to the skills for solving daily problems in a given time and context, and the global scheme which represents the society's knowledge accumulated over the history. These two we respectively model as the payoff and the capacity. \\

\n \textbf{Example.} To clarify our scoring scheme, say in a given iteration $s$ collectivists manage to solve the task in a group $j$ of size $S$. The group's capacity increases by $\chi$, being $\chi$ the number of the individuals in the group which really solved the problem: $\C_j \rightarrow \C_j + \chi$. Fitness updates are as follows: each collectivist, regardless of having solved the task or not, gains $\pi_i \rightarrow \pi_i + \frac{s \C_i}{S}$. Each individualist that solves the task gains $\pi_i \rightarrow \pi_i + \frac{s \C_i}{S} + G_i$. A failing individualist makes the same gain as a collectivist $\pi_i \rightarrow \pi_i + \frac{s \C_i}{S}$. Thus, an individualist can count one the same gain as a collectivist, but if no collectivist solves that task, even that gain is lost. Both capacity and the fitness increase monotonically over time, depending on the distribution of collaboration tendencies $p_i$, and the values of $S$ and $R$. \\

\n \textbf{Rounds and replacement of players.} $n$ groups of size $S$ play the game simultaneously for a given $R$. After 1000 iterations, which we call one round, the game is interrupted. We compute the mean fitness $\bar \pi$ of all players regardless the group they belong to. We remove at random 20 \% of players whose fitness is below $\bar \pi$. They are replaced by new players whose $p_i$ are drawn anew, so that the groups' sizes $S$ are preserved. All groups' capacities $\C_j$ and all players' fitnesses $\pi_i$ are reset to 0. The value $R$ remains the same. Note that the only thing that has changed from one round to another is the structure of groups in terms of players' $p_i$, i.e., the distribution of $p_i$ within each group. Better fit players are kept in the game, in addition to 80 \% of lesser fit players. It is the player's $p_i$ and its relation with other player's $p_i$-s that dictates the player's overall performance in any game. \\

\n \textbf{Full game and full simulation.} A round consists of 1000 iterations, and a game consists of 2000 rounds. Thus, the systems evolves over 2000 rounds, with evolutionary selection being applied at the beginning of each round. We checked that the numbers of iterations and rounds are sufficient for reaching a stable configuration. Upon finishing a game, we are left with the distribution of $p_i$ that optimally fits for solving the task of given simplicity $R$ while operating in a group of size $S$. For better statistics, we conduct 20 games for each choice of $S$ and $R$ and average the results, so that the final values do not depend on random realizations in our model. As we are interested only in the final distribution of $p_i$, the choice of the initial distribution for $p_i$ basically plays no role.

\section*{Results} \label{Results}

For each simulation done for fixed $S$ and $R$, we store the following quantities: $\bar p$, the mean of player's collaboration probabilities $p_i$ in one group, $\bar \pi$, the mean players' fitness $\pi_i$ in one group, and $\Sigma_{max}$, the capacity $\C$ reached by a given group after 2000 rounds (i.e., the final value of $\C$). Each quantity is stored after averaging over 20 games. Our first result, as expected by evolutionary nature of our algorithm, is that for any fixed $R$, the final results for all groups of given size $S$ are the same. So for example, all $n=4$ groups with $S=25$ players ultimately reach the same $\bar p$, $\bar \pi$ and $\C_{max}$ for any given $R$. This means it suffices to consider only one group of size $S$. Therefore, in what follows we represent our findings via final values of $\bar p$, $\bar \pi$ and $\C_{max}$ as function of $R$ and $S$.

\begin{figure}[!ht]  \begin{center} 
       \includegraphics[width=0.6\textwidth]{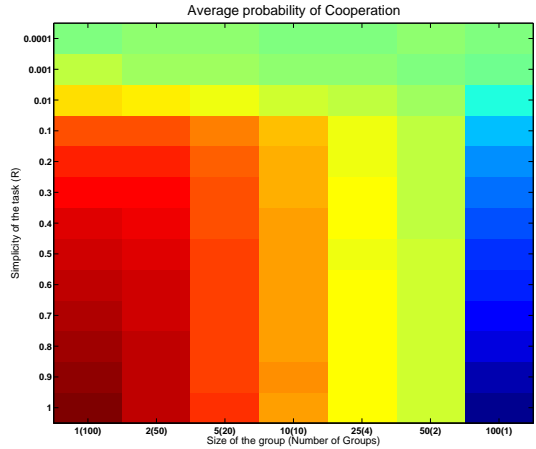}
       \caption{The value of $\bar p$ as a function of group size $S$ and the tasks simplicity $R$. Colors denote $\bar p$ (red - large, green - medium, blue - small). Note that the scale in $R$ is illustrative.} 
  \label{fig-average-p}
\end{center}  \end{figure}

We begin by examining $\bar p$, the mean collaboration probability in a group of size $S$ which is undertaking the task of simplicity $R$. The results are shown in Figure \ref{fig-average-p} as a surfaceplot. For very hard problems (small $R$), players have no preference on collaborating, so $\bar p=0.5$ regardless of $S$. In contrast, for very easy tasks, players in small groups strongly tend to collaborate, while players in big groups tend to work individually.

On the other hand, for small groups the tendency to collaborate increases with increasing task simplicity, while for big groups this tendency actually decreases. In between, for groups of size roughly $S=50$, the collaboration probability is basically constant with $R$, and hence independent on the task that such group is facing.

\begin{figure}[!ht]  \begin{center} 
       \includegraphics[width=\textwidth]{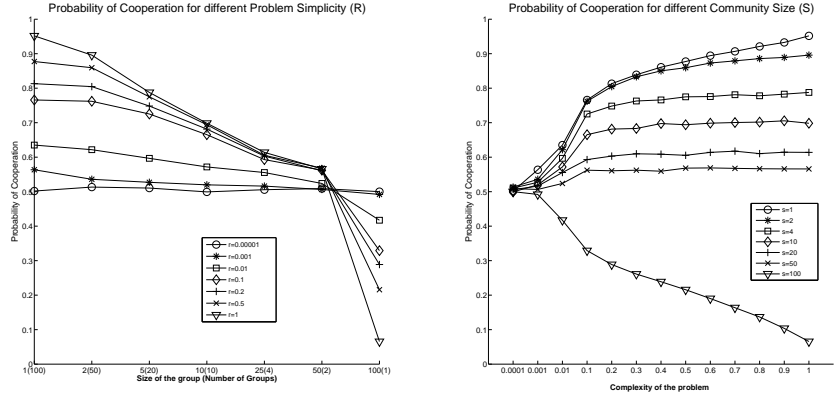}
  \caption{Left: the values of $\bar p$ as function of group size $S$ for varying values of $R$ (see legend). Right: the values of $\bar p$ as function of task simplicity $R$ for varying values of $S$ (see legend).} 
  \label{fig-profiles-for-R-and-S}
\end{center}  \end{figure}

To examine these trends in more detail, we show in Figure \ref{fig-profiles-for-R-and-S} the sequence of profiles of $\bar \pi$ for constant $R$ (left) and for constant $S$ (right).

Looking at the left plot, for the most difficult tasks the $\bar p$ is indeed constant with the group size, while for simpler tasks it exhibits a clear trend: smaller groups tend to collaborate more intensely, while players in the larger groups are predominantly individualists. Groups with approximately 50 players seem always to display $\bar p = 0.5$ regardless of $R$. This is also confirmed by the plots on right in Figure \ref{fig-profiles-for-R-and-S}: all curves with exception of $S=100$ indicate that collaboration propensity increases with task simplicity. This increase becomes more and more moderate with the group size, so that finally, for $S=100$ we find a clear negative trend: collaboration propensity decreases with the task simplicity. In between we find an almost constant curve for roughly $S=50$.

We interpret these findings as follows. When the task is extremely difficult, it is hard to solve it collectively as well as individually. Thus, in such cases there is no particular preference for either approach, regardless of the group size. As the task becomes gradually easier, it is less beneficial for players in small groups to work individually, since this quickly reduces the size of the collaborating subgroup, ultimately making it so small that there is little or no fitness gained from it for anybody. This in turn means that by collaborating a player is more safe to gain at least some fitness. In contrast, easier problems encourage the players in large groups to defect collaboration, since the collaborating subgroup is always fairly large and contributes significant fitness. These circumstances allow for more ``individually oriented'' players to seek their luck by defecting, which they can do safely since at least some fitness is practically guaranteed. This feature captures the concept of free-riders discussed in the Introduction. These players contribute no knowledge of their own to the collective, but at the same time manage to benefit from the novelties reached collectively by the rest of the group. The equilibrium case of groups with approximately $S=50$ players is the \textit{optimal} system where both collaboration and individualism are equally beneficial for task of any difficulty. Such group has a good balance between its collectivist part which slowly but surely generates new knowledge, and its individualists who are still on average equally fit as collectivists.

\begin{figure}[!ht]  \begin{center} 
       \includegraphics[width=\textwidth]{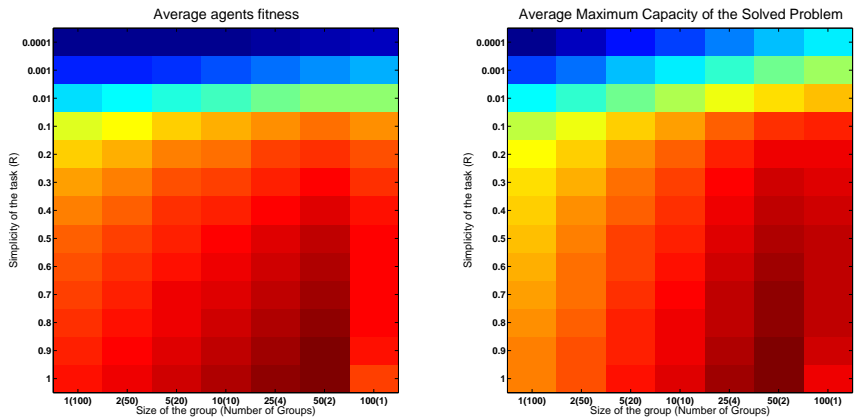}
  \caption{Left: the mean fitness $\bar \pi$ reached by a given group of size $S$ when solving the task of simplicity $R$. Right: the same for maximum capacity $\C_{max}$.} 
  \label{fig-capacity-and-fitness}
\end{center}  \end{figure}

To support these observations with the results for the mean fitness $\bar \pi$ and the maximal capacity $\C_{max}$, we show in Figure \ref{fig-capacity-and-fitness} the surfaceplots for these two quantities. Both surfaceplots display similar features: as expected, the biggest values are found for the simplest tasks when faced by the largest groups.

Smallest $\bar \pi$ and $\C_{max}$ correspond to the most difficult of  tasks when approached by the smallest groups. The group with $S=50$ players optimizes the gain in both quantities, and in fact performs better than the group with $S=100$ players. This resonates with the earlier observation that this group has the best balance between collectivism and individualism. Indeed, by working excessively in isolation and looking too eagerly for free rides, the players in $S=100$ group at the end achieve lesser total fitness, despite their larger size which could have made them all much better fit. It is easy to conclude that this trend would actually further increase with even larger groups, leaving $S=50$ as the only optimal case. This confirms that the knowledge accumulation in too large groups is actually slower than in groups of certain moderate size, when the balanced tendencies for collective and individual approach. On the other hand, one could argue that in a real-life scenario this might in fact be in the long-run beneficial for the society. Namely, the individually gained unique knowledge, that could not have been gained via pure collectivism, might eventually be shared with the rest of the community, and hence available to everyone. By running the simulations with a number of players different from $N=100$, we confirmed that the equilibrium group size is indeed robust to this, and is always approximately $N/2$. \\

\n \textbf{Analytical  approach.} We now show that the above numerical findings can be justified via analytical treatment, relying on standard methods from statistical physics. Let us consider the average fitness for a given player, depending on the player's strategy and the strategy of the opponents. We call $j$ the player's group and $\alpha_j$ the collectivists density in this group. The average fitness of a collectivist in $j$ at time $t$ is then
\be
\langle\pi_c\rangle_j=\alpha_j\left(\C^j_t+1\right) \ , 
\ee
where $\C^j_t$ is the capacity reached at time $t$ in group $j$, and $\langle \cdot \rangle$ indicate averaging. On its turn, individualist (defector) average fitness is
\be
\langle\pi_d\rangle_j=\langle\pi_c\rangle_j+\bar G R^{\bar G} \ , 
\ee
where, by definition, $\bar G=11/2$ is the mean value of the integer random number $G_i\in\{1,2,\dots,10\}$. Of course, within the same group $j$ it is always $\langle\pi_d\rangle_j > \langle\pi_c\rangle_j$, so that in absence of competition between the groups ($i.e.$, $S=N$, $n=1$) individualism is the best strategy, as confirmed by the simulations. On the other hand, for $n \geq 2$, let us think about a group $l$ made up by only collectivists, compared with a group $j$ with just one individualist (we consider this scenario as it the best situation for a defector). In this case, $l$'s collectivist have a better fitness than $j$'s defector if it results
\be
\label{cond1}
\langle\pi_c\rangle_l=\C_t^l+1>\langle\pi_d\rangle_j=\left(1-\frac{1}{S}\right)(\C_t^j+1)+\bar G R^{\bar G}
\ee
Here we distinguish two cases, the large group size $S \gg 1$, and the opposite regime where $S \sim 1$. We treat these two cases separately below.

In the large group size limit, the equation~(\ref{cond1}) reduces to
\be
\label{case1}
\C_t^l+1>(\C_t^j+1)+\bar G R^{\bar G}
\ee
Equation~(\ref{case1}) implies that in order to have the best fitness, collectivist in $l$ must have reached at time $t$ a capacity such that
\be
\label{cond2}
\C_t^l-\C_t^j>\delta=\bar G R^{\bar G} \ . 
\ee
This clearly implies that $\delta$ increases monotonically with $R$, going from 0 to 11/2. This means that, with increasing $R$ cooperative groups have to reach a higher capacity to thrive, but this is balanced by the higher ease to solve the tasks: the combined effect is that, given the group size, the final cooperation level does not depend strongly on $R$, as shown in Figure \ref{fig-average-p}. On the other hand, increasing $R$ enhances the global fitness, because it is easier to solve the tasks and, for a cooperative group, reach a higher capacity with respect to the others, as depicted in equation~(\ref{cond2}). This behavior is confirmed in Figure \ref{fig-capacity-and-fitness}, where an increase for increasing $R$ is observable in the average agent fitness. The narrow tilt change for $R\simeq0.5$ can be understood considering that when $R$ becomes smaller than $R^*=0.5$, also $\delta$ gets much smaller than $\delta^*\simeq0.5$, that is, there is practically no need to reach a higher capacity for cooperators to thrive, lowering the global fitness.

In the limit of $S$ approaching to 1, equation~(\ref{cond1}) reduces to
\be
\label{case2}
\C_t^l>\bar k R^{\bar k} -1 \ ,  
\ee
relation automatically satisfied for $\C_t^l\geq2$, implying that as a group solves the second task, cooperation becomes the best strategy (in other words, free-riding is no longer convenient). This is also confirmed in Figure \ref{fig-average-p}, where it is clear that for small group size cooperation is in general enhanced, and increases more with increasing $R$. This analytical reasoning hence entirely confirms our numerical findings.

\section*{Discussion}  \label{Discussion}

The above numerical and analytical study showed two consistent results. First, when facing the problems whose difficulty is not extreme, small groups tend to collaborate significantly more than large groups. In fact, very large groups end up having lesser average fitness due to excessive individualism being stimulated in these conditions. Small groups are actually known to have weaker knowledge and culture preservation due to not having enough ``critical mass'' of subjects~\cite{derex}. In contrast, large groups have too many free-riders which slow down the knowledge generation process. This agrees with the sociopsychological observations about the sub-population of free-riders, who use the niche provided by the stable knowledge support generated by the collaborating majority. Still, when tasks become more demanding, large groups gradually discourage individualism, and cooperation restores to a level equal to individualism. Second, there exist and optimal group size, where the stimulations for collectivism and individualism are in balance so that both are equally encouraged, regardless of the task difficulty. Note that this effect is exclusively the consequence of the critical group size, and the evolution which keeps the size fixed. Interestingly, groups of exactly this size ultimately manage to have not only the highest fitness, but also the best capacity. This indicates that under proper scenario, optimal utilization of crowdsourcing is possible only by virtue of managing the number of subjects involved in a certain experiment or real-life situation.

The stated conclusion holds even in the presence of defection, which is very realistic. Our evolution was tuned towards optimizing the fitness of individual players, rather than the group's capacity, done by gradually eliminating less fit players. Note however, that under these circumstances, it still holds that the group of critical size performs best in \textit{both} average fitness \textit{and} maximal capacity. That is to say, for the critical size, the system simultaneously reaches both individual and collective optimum, i.e., they do not necessarily exclude each other, as it is usually the case in the game-theoretic models. So, while our results do confirm that free-riding is related to the defection phenomena known from game theory, we show that in the considered model of knowledge dynamics a certain amount of free-riding can play a particular role, as it opens the avenue for individually oriented subjects to gain unique knowledges. As already mentioned, free riding in the real world might ultimately prove advantageous for the society, since this unique knowledge can still at some point be shared with other, and become part of the knowledge support provided for future generations~\cite{uzzi}. Not surprisingly, it is well known that several societies throughout history become powerful by benefiting from inventions coming from global civilization, while withholding their own knowledge and technology generated under these circumstances. Still, such societies in return do eventually provide benefits to the rest of the world via different means~\cite{diamond}.

Limitations of the model involve the fact that players' collaboration probabilities are not affected by outcomes in iterations and rounds. We considered more realistic to adopt an evolutionary model instead, where the group population is adjusted for better performance by searching for the optimal distribution of collaboration probabilities. Another limitation is related to not tuning our model to favor capacity instead of fitness. This would have a different outcome, not allowing for any individualism. However, given the ubiquity of free-riding in nature, we consider more instructive to include it in our model. Of course, there is always an open question of designing better and more realistic models, which with modern computing power is increasingly easy to simulate. As far as robustness to the choice of parameters goes, we have tried several choices of parameters defining our model (for example the span of $G_i$), and have always recovered qualitatively the same results.

The most important direction of this field lies of course in testing the potentials of crowdsourcing via real experiments. This refers to social experiments with controlled rules and known participants, as well as to uncontrolled online experiments, where people are engaged spontaneously and without external organization~\cite{lee}. Although several important results have been recently reported in this direction, a standard framework for undertaking such experiments has yet to be established. Finally, this related to extending the range of applicability of crowdsourcing, chiefly using Internet and tools such as Mechanical Turk~\cite{rand1}, where the crucial issue revolves around findings ways to best stimulate and engage the participants to contribute ideas and thoughts. In conclusion, we hope that our results provide a new stepping stone for better understanding and eventual large-scale usage of crowdsourcing, which is to lead to better exploiting the full potential of our more than ever connected civilization.

\section*{Acknowledgements}  \label{Acknowledgements} 

Work supported by the EU H2020-MSCA-ITN-2015 project COSMOS 642563, by the Creative Core FISNM-3330-13-500033, by the Slovenian Research Agency via program P1-0383 and project J1-5454, Commission (FP7-ICT-2013-10) Proposal No. 611299 SciCafe 2.0., and by project CLARA (CLoud plAtform and smart underground imaging for natural Risk Assessment), funded by the Italian Ministry of Education and
Research (PON 2007-2013: Smart Cities and Communities and Social Innovation; Asse e Obiettivo: Asse II - Azione Integrata per la Societ\`a dell'Informazione; Ambito: Sicurezza del territorio).

\end{document}